\documentstyle[aps]{revtex}
\begin{document}
\title{Fractal Geometry of Normal Phase Clusters and Magnetic Flux Trapping in
High-Tc Superconductors}
\author{Yu. I. Kuzmin}
\address{Ioffe Physical Technical Institute of Russian Academy of Sciences,\\
Polytechnicheskaya 26 St., Saint Petersburg 194021 RUSSIA\\
tel.: +7 812 2479902, fax: +7 812 2471017\\
e-mail: yurk@shuv.ioffe.rssi.ru, iourk@usa.net}
\date{\today}
\maketitle

\begin{abstract}
The effect of geometry and morphology of superconducting structure on
magnetic flux trapping is considered. It is found that the clusters of
normal phase, which act as pinning centers, have significant fractal
properties. The fractal dimension of the boundary of these clusters is
estimated using a simple area-perimeter relation. A superconductor is
treated as a percolation system. It is revealed that the fractality
intensifies the magnetic flux trapping and thereby enhances the critical
current value.
\end{abstract}

One of the important properties of the clusters of a normal phase in
superconductors consists in their capability to act as pinning centers of
high efficiency \cite{k1}-\cite{m5}. When these clusters are out of contact
with one another they can validly trap a magnetic flux, so that the vortices
are held from moving under the action of the Lorentz force. It is the
geometric and morphological properties of such clusters as well as their
effect on the magnetic flux dynamics that will interest us. Let us consider
a superconductor containing inclusions of a normal phase of columnar shape.
When such a superconducting structure is cooled below the critical
temperature in the magnetic field along the direction of the longest size of
those inclusions, the distribution of the trapped magnetic flux will be
two-dimensional. As an example we can consider the superconducting films
where the normal phase inclusions were created during the growth process at
the sites of defects on the boundary with the substrate \cite{m5}-\cite{k7}.
Let us suppose that the film surface fraction covered by the normal phase is
below the percolation threshold for the transfer of magnetic flux (50\% in
the case of 2D-percolation \cite{s8}). At the same time, the relative
portion of superconducting phase exceeds the percolation threshold, so that
the transport current can flow through the superconducting percolation
cluster. Such a structure provides for effective pinning and thereby raises
the critical current, because the magnetic flux is trapped in the finite
normal phase clusters. When the transport current is rising, the trapped
magnetic flux will be changed, for the vortices will break away from the
clusters of pinning force weaker than the Lorentz force created by the
current. Since the infinite superconducting cluster is formed in the film,
the vortices must cross a region occupied by the superconducting phase when
they are coming out from the pinning centers.

In the subsequent examination it must be taken into account that high-Tc
superconductors (HTSC) are characterized by an extremely short coherence
length, so weak links form readily in these materials. Various structural
defects as grain boundaries, crystallites, and especially twins, which would
simply cause some additional scattering at long coherence length, give rise
to the weak links over a wide range of scales in HTSC \cite{h2}, \cite{m5}, 
\cite{den9}. Moreover, the presence of a magnetic field is favorable to the
formation of weak links, because its action further reduces a coherence
length \cite{s10}. Thus around each normal phase cluster there are always
weak links through which the vortices may pass under the action of the
Lorentz force. The larger the cluster size, the more such weak sites are
along its border with the surrounding superconducting space, and therefore,
the smaller is the critical current at which the magnetic flux breaks away
from a given pinning center. It appears that each normal phase cluster has
its own value of the critical current, which make the contribution to the
total statistical distribution. On the basis of these geometric arguments,
let us suppose that this critical current $I$ is inversely proportional to
the perimeter $P$ of the normal phase cluster: $I\propto 1/P$ , because the
larger the perimeter is, the higher is the probability to get a weak link
there. At this point we shall assume that the weak links concentration per
unit perimeter length is constant for all clusters and that all the clusters
of equal area have the same pinning force.

On the other hand, when a transport current is slowly increased, magnetic
flux will break away first from the clusters of small pinning force, of
small critical current, and therefore, of large size. Thus the decrease in
the trapped magnetic flux $\Delta \Phi $ caused by a transport current is
proportional to the number of all the normal phase clusters of critical
currents less than a preset value. Taking into account that the magnetic
flux trapped into a single cluster is proportional to its area $A$, the
relative decrease in the total trapped flux can be expressed with the
accumulated probability function $F=F(A)$ for the distribution of the areas
of the normal phase clusters, which is a measure of the number of the
clusters of area smaller than a given value of $A$: 
\begin{equation}
\frac{\Delta \Phi }{\Phi }=1-F(A)\text{ \ \ \ \ \ \ \ \ \ , \ \ \ \ \ \ \ \
where \ \ \ \ \ \ \ \ \ \ }F(A)={\sf P}\{\forall A_{j}<A\}  \label{dist1}
\end{equation}
Here ${\sf P}$ is the probability that any $j$th cluster has the area $A_{j}$
less than a given value $A$.

Thus, in order to clear up how the transport current acts on the trapped
magnetic flux, it is necessary to find out the relationship between the
perimeter of the normal phase clusters and the distribution of their areas.
To analyze the geometric probability characteristics of normal phase
clusters an electron photomicrograph of YBCO film, which was similar to that
published earlier in Ref.~\cite{k7}, has been studied. The region of a total
area of 200\thinspace $\mu $m$^{2}$ containing 528 clusters has been
scanned. The areas and perimeters of the clusters have been measured by
covering their digitized pictures with a rectangular grid of resolution cell 
$60\times 60\,$nm$^{2}$. The results of the statistical treatment of those
data are presented in the Table~\ref{table1}. The normal phase occupies 20\%
of the total area only, so the percolation superconducting cluster is dense
enough. A high skewness (1.76) as well as the statistically insignificant
(5\%) difference between the sample mean and the standard deviation allow us
to conclude that there is an exponential distribution for areas of the
normal phase clusters which has such an accumulated probability function 
\begin{equation}
F(A)=1-\exp \left( -\frac{A}{\overline{A}}\right)  \label{ac2}
\end{equation}
where $\overline{A}$ is the mean area of the cluster.

Furthermore, it might be natural to believe that the perimeter-area relation
for the normal phase clusters obeys the well known geometric formula: $%
P\propto \sqrt{A}$. However, it would be a very rough approximation only,
because this relationship holds for Euclidean geometric objects but on no
account for the fractals. It is just the fractal nature of the normal phase
clusters that is of fundamental importance here. For fractal geometric
objects perimeter-area relation must be re-written in a different way \cite
{m11}: 
\begin{equation}
P\propto A^{\frac{D}{2}}  \label{scaling3}
\end{equation}
where $D$ is the fractal dimension of the perimeter. This is consistent with
the generalized Euclid theorem \cite{m12}, which states that, for similar
geometric objects, the ratios of corresponding measures are equal when
reduced to the same dimension. Hence it follows that $P^{1/D}\propto A^{1/2}$%
, which is applicable both to fractal ($D>1$) and Euclidean ($D=1$) figures.

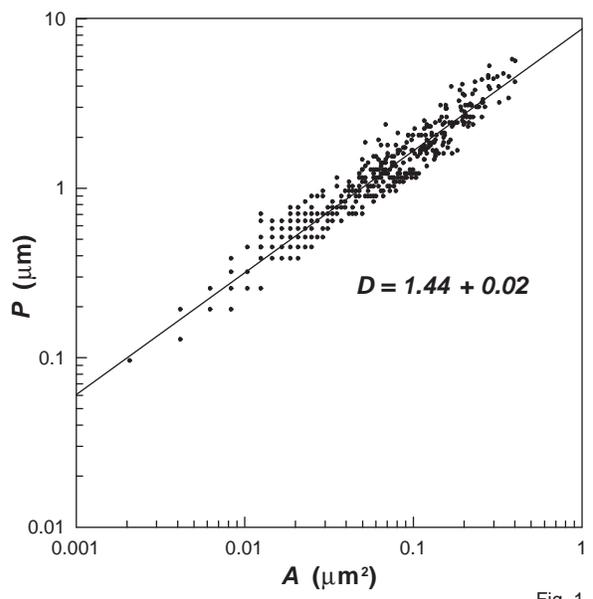
\begin{figure}[tbp]
\caption{Perimeter-area relationship for the normal phase clusters. The
results of photomicrograph scanning are shown by points; the line indicates
the least squares regression.}
\label{figure1}
\end{figure}

The perimeter-area relation for the normal phase clusters is shown in Fig.~%
\ref{figure1}. Although all 528 points mentioned above are plotted on the
graph, certain of them coincide for the finite resolution of the picture
digitization. This feature is noticeable, primarily, for small clusters:
some points at the lower left of Fig.~\ref{figure1} are arranged discretely
with the spacing equal to the limit of resolution (60\thinspace nm). 
Fig.~\ref{figure1} demonstrates such an important peculiarity:
perimeter-area relation obeys the scaling law of Eq.~(\ref{scaling3}). There
are no apparent crossovers or breaks on this graph over the range of more
than two orders of magnitude in area. These data apply to the clusters of
different shape, nevertheless all the points fall on a straight line in
logarithmic scale. Thus, the fractal dimension $D$ of the cluster boundary
can be found from the slope of the regression line of the form of Eq.~(\ref
{scaling3}); a least squares treatment of the perimeter-area data gives the
estimate of $D=1.44\pm 0.02$ with correlation coefficient 0.93. The scaling
perimeter-area behavior implies an absence of any characteristic length
scale between 0.1\thinspace $\mu $m and 10\thinspace $\mu $m in the linear
size of the normal phase cluster. Hence these clusters have all the
properties of stochastic fractals \cite{m12}.

The problem on the fractal cluster perimeter is quite similar to the one on
the fractal boun\-dary of clouds in atmosphere \cite{l13}, \cite{r14}. The
main distinction is that instead of the projection of a cloud on the Earth's
surface we are considering the cross-section of the extended columnar object
by the plane carrying a transport current. Therefore, though a normal phase
cluster, as well as a cloud, represents a self-affine fractal \cite{m15}, it
is possible to consider its geometric probability properties in the planar
section only, where the boundary of the cluster is statistically
self-similar.

Next, using relation of Eq.~(\ref{scaling3}) between the fractal perimeter
and the area of the cluster, for critical current we get: $I=\alpha A^{-D/2}$%
, where $\alpha $ is the form factor. In accordance with starting formula (%
\ref{dist1}), the exponential distribution of cluster area of Eq.~(\ref{ac2}%
) gives rise to an exponential-hyperbolic distribution of critical currents: 
\begin{equation}
\frac{\Delta \Phi }{\Phi }=\exp \left[ -\left( \frac{2+D}{2}\right) ^{\frac{2%
}{D}+1}i^{-\frac{2}{D}}\right]  \label{exhyp4}
\end{equation}
where $i\equiv I/I_{c}$ , and $I_{c}=\left( 2/\left( 2+D\right) \right)
^{\left( 2+D\right) /2}\alpha \left( \overline{A}\right) ^{-D/2}$ is the
critical current of the resistive transition.

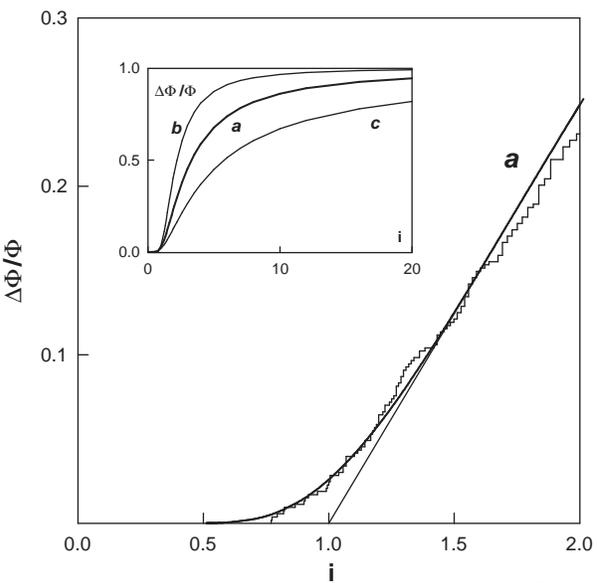
\begin{figure}[tbp]
\caption{Effect of a transport current on trapped magnetic flux. On the main
chart the thick line (a) shows the decrease in trapped flux for the fractal
dimension $D=1.44$; thin straight line is the tangent drawn through the
inflection point of the previous curve; step line is the empirical
distribution function of the critical currents. In the inset line (a)
corresponds to the case of $D=1.44$; line (b) to $D=1$; line (c) to $D=2$.}
\label{figure2}
\end{figure}

Thus, the entire distribution of critical currents determined by the
geometric probability properties of the normal phase clusters can be found
with the aid of Eq.~(\ref{exhyp4}) from the changes in the trapped magnetic
flux caused by transport currents of different amplitude. The effect of the
transport current on the trapped flux is illustrated in Fig.~\ref{figure2}.
The critical current distribution of Eq.~(\ref{exhyp4}) for fractal
dimension $D=1.44$ is drawn there by the line (a). 

In order to get the relationship between the dynamics of the trapped
magnetic flux and the geometric morphological properties of the
superconducting structure the empirical distribution function $F^{\ast
}=F^{\ast }\left( A\right) $ for the sampling of the areas of normal phase
clusters has been computed (step line in Fig.~\ref{figure2}). This function
gives a statistical estimate of the accumulated probability function $%
F=F\left( A\right) $. The value $F^{\ast }\left( A\right) $ was calculated
for each order statistic as the relative number of clusters of area smaller
than a given value $A$. A co-ordinate transformation of the form 
\[
\left\{ F^{\ast }\rightarrow 1-F^{\ast }\text{ \ \ , \ \ \ \ \ \ }%
A\rightarrow \left( \frac{2+D}{2}\right) ^{\frac{2+D}{2}}\left( \frac{%
\overline{A}}{A}\right) ^{\frac{D}{2}}\right\} 
\]
allows us to get an empirical function of distribution of the normal phase
clusters in their critical currents, which is a statistical analog of the
distribution function of Eq.~(\ref{exhyp4}). This distribution reflects the
morphological properties of the superconducting structure of the specimen,
and as it is seen from the Fig.~\ref{figure2}, both distributions coincide
well (the thick line (a) and the step one on the main chart).

Furthermore, Fig.~\ref{figure2} shows that currents of amplitude less than $%
i_{\min }=0.77$ have no action upon the trapped flux because there are no
pinning centers of such small critical currents in the specimen. Since the
pinning force is smaller for larger clusters, the constancy of the trapped
flux in this range of currents is attributed to the absence of such normal
phase clusters that have the perimeter greater than a certain threshold
value. The magnitude of the corresponding critical current can be obtained
from the formula 
\[
i_{\min }=\left( \frac{2+D}{2}\right) ^{\frac{2+D}{2}}\left( \frac{\overline{%
A}}{A_{\max }}\right) ^{\frac{D}{2}} 
\]
where $A_{\max }$ is the area of the largest cluster in the sampling (see
``Max value'' in the Table~\ref{table1}).

It is important to note that the specimen may remain superconducting under
the action of pulsed current of amplitude even greater than the critical
current of resistive transition ($i=1$), because the thermo-magnetic
instability, which would be inevitable when a constant current of the same
amplitude flowed through the specimen, does not manage to develop during a
short-lived disturbance.

The inset on Fig.~\ref{figure2} shows how the fractal dimension affects the
trapping of the magnetic flux. The curve of $D=1$ (line (b)) relates to
usual Euclidean clusters; the curve of $D=2$ (line (c)) applies to the case
of the greatest possible fractal dimension for curves in a plane (as an
example, the plane-filling Hilbert curve of infinite order has such a
fractal dimension). Whatever the geometric morphological properties of the
normal phase clusters may be, the graph for their transport current
dependence of the trapped magnetic flux will fall within the region bounded
by those two limiting curves (like line (a)). The practically important
conclusion follows from Fig.~\ref{figure2}: fractality of the normal phase
clusters strengthens the trapping of the flux, thus hindering the vortices
to break away from the pinning centers, thereby the current-carrying
capability of superconductor with such fractal clusters is enhanced.
Actually, the transport current of a magnitude $i=2$ causes the 43\% of the
total trapped flux to break away from the usual Euclidean clusters ($D=1$%
\thinspace , line (b)), whereas this value is equal only to 13\% for the
normal phase clusters of a maximal fractality ($D=2$\thinspace , line (c))
what is equivalent that the pinning is three times greater in the latter
case.

Thus, in the present work the fractal nature of the normal phase clusters is
found and its effect on the magnetic flux trapping in superconductors is
revealed. The fractal dimension of the cluster perimeter is estimated. It is
shown that the fractality contributes to the trapping of the magnetic flux
that can cause the critical current to increase.

%TCIMACRO{
%\TeXButton{B}{\begin{table}[tbp] \centering%
%}}%
%BeginExpansion
\begin{table}[tbp] \centering%
%
%EndExpansion
\caption{Statistics of normal phase clusters\label{table1}} 
\begin{tabular}{ccc}
& Area in $\mu $m$^{2}$ & Perimeter in $\mu $m \\ \hline
Mean & $0.0765$ & $1.293$ \\ 
Sample standard deviation & $0.0726$ & $0.962$ \\ 
Standard error of estimate & $3.16\times 10^{-3}$ & $0.0419$ \\ 
Total scanned amount & $40.415$ & $682.87$ \\ 
Min value & $2.07\times 10^{-3}$ & $0.096$ \\ 
Max value & $0.4015$ & $5.791$ \\ 
Skewness & $1.76$ & $1.67$%
\end{tabular}
%TCIMACRO{
%\TeXButton{E}{\end{table}%
%}}%
%BeginExpansion
\end{table}%
%
%EndExpansion

\end{document}